\documentclass[journal]{IEEEtran}
\IEEEoverridecommandlockouts
\usepackage{amsmath}
\usepackage{graphicx}
\usepackage{amssymb}
\usepackage{cases}
\usepackage[ruled,vlined,lined,ruled,linesnumbered]{algorithm2e}
\usepackage{cite}
\usepackage{fancyhdr}
\usepackage{relsize}
\usepackage{float}
\usepackage{longtable}
\usepackage{nomencl}
\makenomenclature

\begin{document}
\title{Graphical Methods for Defense Against False-data Injection Attacks on Power System State Estimation}
\author{Suzhi~Bi,~\IEEEmembership{Member,~IEEE} and Ying Jun (Angela) Zhang,~\IEEEmembership{Senior Member,~IEEE}
        \thanks{This work was supported in part by the National Natural Science Foundation of China (Project number 61201261), the National Basic Research Program (973 program Program number 61101132) and the Competitive Earmarked Research Grant (Project Number $419509$) established under the University Grant Committee of Hong Kong.}
        \thanks{S.~Bi is with the Department of Electrical and Computer Engineering, National University of Singapore, Singapore 117583. (Email: bsz@nus.edu.sg).}
        \thanks{Y.~J.~Zhang is with the Department of Information Engineering, The Chinese University of Hong Kong, Shatin, New Territories, Hong Kong, and Shenzhen Research Institute, The Chinese University of Hong Kong, Shenzhen, China 518057. (Email: yjzhang@ie.cuhk.edu.hk).}}
\maketitle

\begin{abstract}
The normal operation of power system relies on accurate state estimation that faithfully reflects the physical aspects of the electrical power grids. However, recent research shows that carefully synthesized false-data injection attacks can bypass the security system and introduce arbitrary errors to state estimates. In this paper, we use graphical methods to study defending mechanisms against false-data injection attacks on power system state estimation. By securing carefully selected meter measurements, no false data injection attack can be launched to compromise any set of state variables. We characterize the optimal protection problem, which protects the state variables with minimum number of measurements, as a variant Steiner tree problem in a graph. Based on the graphical characterization, we propose both exact and reduced-complexity approximation algorithms. In particular, we show that the proposed tree-pruning based approximation algorithm significantly reduces computational complexity, while yielding negligible performance degradation compared with the optimal algorithms. The advantageous performance of the proposed defending mechanisms is verified in IEEE standard power system testcases.
\end{abstract}


\section{Introduction}
\subsection{Motivations and summary of contributions}
\IEEEPARstart{T}{he} current power systems are continuously monitored and controlled by EMS/SCADA (Energy Management System and Supervisory Control and Data Acquisition) systems in order to maintain the operating conditions in a normal and secure state \cite{2004:Abur}. In particular, the SCADA host at the control center processes the received meter measurements using a state estimator, which filters the incorrect data and derives the optimal estimate of the system states. These state estimates will then be passed on to all the EMS application functions, such as optimal power flow, etc, to control the physical aspects of the electrical power grids.

However, the integrity of state estimation is under mounting threat as we gradually transform the current electricity infrastructures to future smart power grids, which are more open to the outside networks from the extensive use of internet-based protocols in the communication system. In particular, enterprise networks and even individual users are allowed to connect to the power network information infrastructure to facilitate data sharing \cite{2008:Ten}. With these entry points introduced to the power system, potential complex and collaborating malicious attacks are brought in as well. Liu \emph{et al}.\cite{2009:Liu} showed that a new false-data injection attack could circumvent bad data detection (BDD) in today's SCADA system and introduce arbitrary errors to state estimates without being detected. Such an attack is referred to as an undetectable false-data injection attack. A recent experiment in \cite{2011:Teixeira} demonstrates that the attack can cause a state-of-the-art EMS/SCADA state estimator to produce a bias of more than $50\%$ of the nominal value without triggering the BDD alarm. Biased estimates could directly lead to serious social and economical consequences. For instance, \cite{2012:Choi,2012:Jia,2011:Xie} showed that attackers equipped with data injection can manipulate the electricity price in power market. Worse still, \cite{2011:Yuan} warned that the attack can even cause regional blackout.

Being aware of its imminent threats to power system, a number of studies are devoted to both understanding its attacking patterns and providing effective countermeasures \cite{2013:Giani,2010:Dan,2012:Cui}. A common approach to mitigate false-data injection attack is to secure meter measurements by, for example, guards, video monitoring, or tamper-proof communication systems, to evade malicious injections \cite{2010:Bobba,2012:Vukovic,2012:Kim}. Recent studies have proposed a number of methods to select meter measurements for protection. For instance, \cite{2010:Bobba} proved that it is necessary and sufficient to protect a set of \emph{basic measurements} so that no undetectable false-data injection attack can be launched. However, the protection scheme in \cite{2010:Bobba} is costly in that the size of a set of \emph{basic measurements} is the same as the number of unknown state variables in the state estimation problem, which could be up to several hundred in a large-scale power system. Under limited budget, the system operator should protect a subset of state variables. This is because an ill-advised protection method may leave the attackers the chance to formulate undetectable attack to compromise a large number of, if not all the state variables, even if many measurements have been secured \cite{2011:Bi}. In this case, the system operator may give priority to protecting the state variables that have greater social/economic impact once compromised, such as those for critical buses/substations connected to heavily loaded or economically important areas, or with critical interconnection purposes \cite{2008:Chakrabarti,2001:NERC,2004:NERC}. On the other hand, even if the system operator has enough budget to defend all the state variables, protecting a set of basic measurements in a random sequence may still open to attackers the possibility to compromise a large number of state variables during the lengthy security installation period. In both cases, it is valuable to devise a method that gives priority to defending a subset of state variables that serves our best interests at the current stage, and opens to the possibility of expanding the set of protected state variables in the future.

In this paper, we focus on using graphical methods to derive efficient strategies that defend any subset of state variables with minimum number of secure measurements. Our detailed contributions are listed as follows,
\begin{itemize}
  \item We derive conditions to select a set of meter measurements, so that no undetectable attack can be launched to compromise a given set of state variables if the selected meters are secured. The conditions are particularly useful in formulating the optimal protection problem that defends the state variables with a minimum cost.
  \item We characterize the optimal protection problem as a variant Steiner tree problem in a graph. Then, two exact solution methods are proposed, including a Steiner vertex enumeration algorithm and a mixed integer linear programming (MILP) formulation derived from a network flow model. In particular, the proposed MILP formulation reduces the computational complexity by exploiting the graphical structure of the optimal solution.
  \item To tackle the intractability of the problem, we also propose a polynomial-time tree-pruning heuristic (TPH) algorithm. With a proper parameter, simulation results show that it yields close-to-optimal solution, while significantly reducing the computational complexity. For instance, the TPH solves a problem of a $300$-bus testcase in seconds, which may take days by the MILP formulation.
\end{itemize}
The proposed MILP and TPH algorithms can also be extended to achieve incremental protection. That is, starting from a set of protected state variables and measurements, the method can gradually expand the set of protected state variables until the entire set of state estimates is protected. The incremental protection method can be used to plan a long-term security upgrade project in a large-scale power system.

\subsection{Related works}
State estimation protection is closely related to the concept of power network observability. The conventional power network observability analysis studies whether a unique estimate of all unknown state variables can be determined from the measurements \cite{2004:Abur}. From the attacker's perspective, \cite{2009:Liu} proved that an undetectable attack can be formulated if removing the measurements it compromises will make the power system unobservable. Conversely, \cite{2010:Bobba} showed that no undetectable attack can be formulated if the power system is observable from the protected meter measurements. In this paper, we extend the conventional wisdom of power network observability to a generalized \emph{state variable observability} to study the protection mechanisms for any set of state variables.

Graphical method is commonly used for power system observability analysis. The early work by Krumpholz \emph{et al}.\cite{1980:Krumpholz} stated that a power system is observable if and only if it contains a spanning tree that satisfies certain measurement-to-transmission-line mapping rules. A follow-up work presented a max-flow method to find such mapping to examine the observability of a power network \cite{1986:Barglela}. Few recent papers also applied graphical methods to study the attack/defending mechanisms of false-data injection. For instance, based on the results in \cite{1980:Krumpholz}, \cite{2011:Kosut} proposed an algorithm to quantify the minimum-effort undetectable attack, i.e. the non-trivial attack that compromises least number of meters without being detected. Besides, \cite{2011:Sou} used a min-cut relaxation method to calculate the security indices defined in \cite{2010:Sandberg} to quantify the resistance of meter measurements in the presence of injection attack. Similar min-cut approach was also applied in \cite{2012:Sou} to identify the critical points in the measurement set, the loss of which would render the power system unobservable.

The problem of defending a subset of critical state variables against undetectable attack was first studied in our earlier work \cite{2011:Bi}, where we proposed an arithmetic greedy algorithm which finds the minimum set of protected meter measurements by gradually expanding the set of secure state variables. However, the computational complexity of the greedy algorithm can be prohibitively high in large scale power systems. For instance, it may take years to obtain a solution in a $57$-bus system. In contrast, we study in this paper the optimal protection from a graphical perspective. By exploiting the graphical structures of the optimal solution, the proposed MILP formulation obtains the optimal solution with significantly reduced complexity. In addition, we also propose a pruning-based heuristic that yields near-optimal solutions in polynomial time.

The rest of this paper is organized as follows. In Section II, we introduce some preliminaries about state estimation and false-data injection attack. We characterize the optimal protection problem in a graph in Section III and propose solution algorithms in Section IV. In Section V, we discuss the methods to extend the proposed algorithms to some practical scenarios, including the method to achieve incremental protection. Simulation results are presented in Section VI. Finally, the paper is concluded in Section VII.

\section{Preliminary}
\subsection{DC measurement model and state estimation}
\begin{figure}
\centering
  \begin{center}
    \includegraphics[width=2.8in]{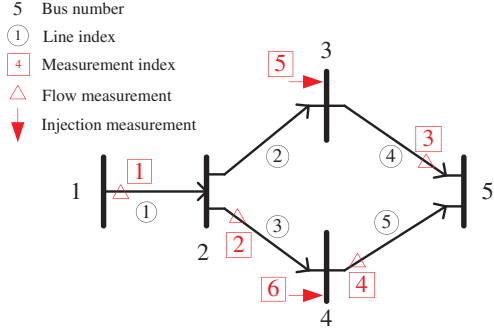}
  \end{center}
  \caption{Measurement placement of an example 5-bus system.}
  \label{61}
\end{figure}

We consider the linearized power network state estimation problem in a steady-state power system with $n+1$ buses. The states of the power system include the bus voltage phase angles and voltage magnitudes. The voltage magnitudes can often be directly measured, while the values of phase angles need to be obtained from state estimation \cite{1994:Grainger}. In the linearized (DC) measurement model, we assume the knowledge of voltage magnitudes at all buses ($1$ in the per-unit system) and estimate the phase angles based on the active power measurements, i.e. the active power flows along the power lines and active power injections at buses \cite{2004:Abur}. By choosing an arbitrary bus as the reference with zero phase angle, the network state consisting of the $n$ unknown voltage phase angles is captured in a vector $\boldsymbol{\theta}=\left(\theta_1,\theta_2,..,\theta_n\right)'$. In the DC measurement model, the $m$ received measurements $\mathbf{z}=\left(z_1,z_2,..,z_m\right)'$ are related to the network states as
\begin{equation}
\label{18}
\mathbf{z} = \mathbf{H}\boldsymbol{\theta} + \mathbf{e}.
\end{equation}
Here, $\mathbf{H}$ is the measurement Jacobian matrix \cite{2004:Abur}. $\mathbf{e}\thicksim \mathcal{N}\left(\mathbf{0},\mathbf{R}\right)$ is independent measurement noise with covariance $\mathbf{R}$. Using a $5$-bus power system in Fig. $\ref{61}$ for example. By setting bus $1$ as the reference bus, there are four unknown state variables. Suppose that the reactance of all transmission lines equals $1$, the measurement Jacobian matrix is
\begin{equation}
\mathbf{H}=\left(
  \begin{array}{ccccc}
    -1 & 0  & 0  & 0\\
    1 & 0 & -1  & 0\\
    0 & -1  & 0  & 1\\
    0 &  0  & 1  & -1\\
    -1 & 2  & 0 & -1\\
    -1 & 0  &  2 & -1\\
  \end{array}
\right),
\end{equation}
where the first $4$ rows correspond to flow measurements while the last two rows correspond to injection measurements. The four columns correspond to bus $2$ to $5$, respectively. Notice that the column corresponding to the reference bus is not included.

When $\mathbf{H}$ is full column rank, i.e. $rank\left(\mathbf{H}\right)=n$, the maximum likelihood estimate $\boldsymbol{\hat{\theta}}$ is given by
\begin{equation}
\label{3}
\boldsymbol{\hat{\theta}} = \left(\mathbf{H}^{T}\mathbf{R}^{-1}\mathbf{H}\right)^{-1}\mathbf{H}^{T}\mathbf{R}^{-1}\mathbf{z} \triangleq \mathbf{Pz}.
\end{equation}
Since $rank\left(\mathbf{H}\right)\leq m$, i.e. the number of rows in $\mathbf{H}$, at least $n$ meters are needed to derive a unique state estimation. Meanwhile, the other $m-n$ measurements provide the redundancy to improve the resistance against random errors.

Errors could be introduced due to various reasons, such as device misconfiguration and malicious attacks. The current power systems use BDD mechanism to remove the bad data assuming that the errors are random and unstructured. It calculates the residual $\mathbf{r}=\mathbf{z}-\mathbf{H\boldsymbol{\hat{\theta}}}$ and compares its $l_2$-norm with a prescribed threshold $\tau$. A measurement $\mathbf{z}$ is identified as a bad data measurement if
\begin{equation}
\label{4}
r=||\mathbf{z}-\mathbf{H\boldsymbol{\hat{\theta}}}||= ||\left(\mathbf{I - \mathbf{HP}}\right) \mathbf{e}||> \tau,
\end{equation}
where $\mathbf{I}$ is an identity matrix. Otherwise, $\mathbf{z}$ is considered as a normal measurement.

\subsection{Undetectable attacks and protection model}
Suppose that attackers inject malicious data $\mathbf{a}=\left(a_1,a_2,..,a_m\right)'$ into measurements. Then, the received measurements become
\begin{equation}
\label{5}
\mathbf{\tilde{z}} = \mathbf{H}\boldsymbol{\theta} + \mathbf{e} +\mathbf{a}.
\end{equation}
In general, $\mathbf{a}$ is likely to be identified by the BDD if it is unstructured. Nevertheless, it is found in \cite{2009:Liu} that some well-structured injections, such as those with $\mathbf{a}=\mathbf{Hc}$, can bypass BDD. Here $\mathbf{c} = \left(c_1,c_2,..,c_n\right)'$ is a random vector. This can be verified by calculating the residual in (\ref{5}), where
\begin{equation}
\tilde{r} = ||\mathbf{\tilde{z}}- \mathbf{HP\tilde{z}} || = ||\mathbf{z}+\mathbf{a}-\mathbf{H}(\mathbf{\boldsymbol{\hat{\theta}}+c})||= ||\mathbf{z}-\mathbf{H\boldsymbol{\hat{\theta}}}||.
\end{equation}
The same residual is obtained as if no malicious data were injected. Therefore, a structured attack $\mathbf{a}=\mathbf{Hc}$ will not be detected by BDD. In this case, the system operator would mistake $\boldsymbol{\hat{\theta}}+\mathbf{c}$ for a valid estimate, and thus an error vector $\mathbf{c}$ has been introduced without being detected.

The risks of undetectable attacks can be mitigated if the system operator can secure measurements to evade malicious injections. Within this context, we assume that the system operator's objective is to ensure that no undetectable attack can be formulated to compromise a given set of state variables $\mathcal{D}\subseteq \mathcal{I}$, where $\mathcal{I}$ is the set of all unknown state estimates. That is, $c_i=0$ for all $i\in \mathcal{D}$. This is achieved by securing a set of meter measurements $\mathcal{P}\subseteq \mathcal{M}$, where $\mathcal{M}$ is the set of all the meters. In other words, attackers are not able to inject false data to any protected meter measurement, i.e. $a_i = 0$, $\forall i\in \mathcal{P}$.

From \cite{2011:Bi}, securing a set of meters $\mathcal{P}$ would eliminate the possibility of undetectable attack to compromise a set of state variables $\mathcal{D}$, if and only if
\begin{equation}
\label{11}
rank\left(\mathbf{H}_{\{\mathcal{P}\},*}\right) = rank \left(\mathbf{H}_{\{\mathcal{P}\},\{\mathcal{I}\setminus \mathcal{D}\}}\right) + |\mathcal{D}|.
\end{equation}
Here, $\mathbf{H}_{\{\mathcal{P}\},*}$ is the submatrix of $\mathbf{H}$ including the rows that correspond to $\mathcal{P}$ and $\mathbf{H}_{\{\mathcal{P}\},\{\mathcal{I}\setminus \mathcal{D}\}}$ is the submatrix of $\mathbf{H}_{\{\mathcal{P}\},*}$ excluding the columns that correspond to $\mathcal{D}$. $|\mathcal{D}|$ denotes the size of $\mathcal{D}$. Naturally, we are interested in minimizing the cost to protect the state variables $\mathcal{D}$. For simplicity, we assume a fixed cost, e.g. manpower or surveillance installation cost, of securing each meter for the time being. This requires solving the following problem
\begin{equation}
\label{98}
\begin{aligned}
& \underset{\mathcal{P}\subseteq\mathcal{M}}{\text{minimize}} & &  |\mathcal{P}|\\
& \text{subject to} & & rank\left(\mathbf{H}_{\{\mathcal{P}\},*}\right)= rank \left(\mathbf{H}_{\{\mathcal{P}\},\{\mathcal{I}\setminus \mathcal{D}\}}\right) + |\mathcal{D}|,\\
\end{aligned}
\end{equation}
which is proved to be an \emph{NP-hard} problem in the next section.

\section{Graphical Characterizations of Optimal State Variable Protection}
Interestingly, we show that (\ref{98}) can be characterized as a variant Steiner tree problem in a graph. The results will be used in the next section to develop graphical algorithms.

\subsection{Network observability and state variable protection}
In this subsection, we first introduce some definitions to characterize a power network in a graph. Then, we establish the equivalence between power network observability and state estimate protection criterion. The results will be used in the next subsection to formulate an equivalent graphical characterization of the optimal state protection problem in (\ref{98}).

A power network can be described in an undirected graph, where vertices and edges represent buses and transmission lines, respectively. We use $e^{(h)}_i$ and $e^{(t)}_i$ to denote the two vertices connected to the edge $e_i$, and $\mathcal{N}_j$ to denote the set of edges incident to vertex $v_j$. The following Definition $1$ gives the notion of measurability in a power network.

\textbf{Definition $1$: (measurability)} The \emph{measured subnetwork} of a meter $r$, denoted by $\bar{G}\left(r\right)$, consists of the vertices and edges \emph{measured} by the meter $r$. That is, for a flow meter $r$ on transmission line $e_i$, $\bar{G}\left(r\right)$ includes the two vertices $\left\{e^{(h)}_i,e^{(t)}_i\right\}$ and edge $e_i$. For an injection meter $r$ at bus $v_j$, $\bar{G}\left(r\right)$ includes the vertex set $\left\{e^{(h)}_i,e^{(t)}_i \mid e_i\in \mathcal{N}_j\right\}$ and edge set $\left\{e_i\mid e_i\in \mathcal{N}_j\right\}$. The \emph{measured subnetwork} of a set of meters $\mathcal{\bar{M}}\subseteq \mathcal{M}$ is defined as
\begin{equation}
\bar{G} (\mathcal{\bar{M}}):=\bigcup_{r\in \mathcal{\bar{M}}}\bar{G}(r).
\end{equation}
In particular, $\bar{G}\left(\mathcal{M}\right)$ is referred to as the \emph{measured full network}.

\begin{figure}
\centering
  \begin{center}
    \includegraphics[width=3.3in]{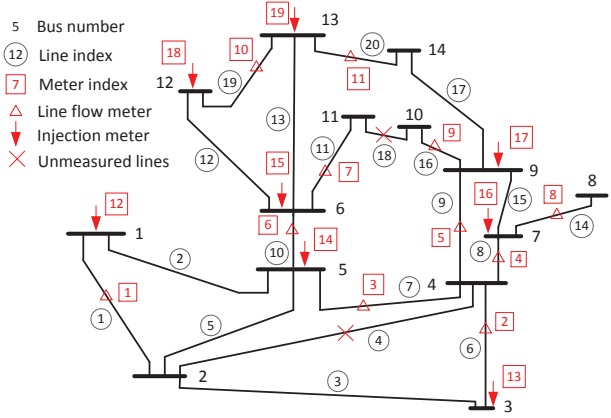}
  \end{center}
  \caption{A measurement placement for the IEEE 14-bus testcase.}
  \label{65}
\end{figure}

Using a $14$-bus testcase in Fig. $\ref{65}$ for example. The measured subnetwork of the flow meter $r_6$ includes edge $e_{10}$ and vertices $v_5$ and $v_6$, i.e.
\begin{equation}
\bar{G}\left(r_6\right) = \left(\left\{v_5,v_6\right\},\left\{e_{10}\right\}\right).
\end{equation}
The measured subnetwork of the injection meter $r_{12}$ is
\begin{equation}
\bar{G}\left(r_{12}\right) = \left(\left\{v_1,v_2,v_5\right\},\left\{e_{1},e_2\right\}\right).
\end{equation}
Besides, the measured subnetwork of $\mathcal{\bar{M}}=\left\{r_6,r_{12}\right\}$ is
\begin{equation}
\bar{G}\left(\mathcal{\bar{M}}\right) = \left(\left\{v_1,v_2,v_5,v_6\right\},\left\{e_1,e_2,e_{10}\right\}\right).
\end{equation}

The conventional power network observability analysis studies whether a unique estimate of all unknown state variables can be determined\cite{2004:Abur}. Here, we extend the concept of network observability to a generalized state variable observability in the following Definition $2$. With a bit abuse of notation, we use a set of vertices $\mathcal{V'}$ to denote the corresponding state variables.

\textbf{Definition $2$: (observability)} A set of state variables $\mathcal{D}\subseteq \mathcal{I}$ is \emph{observable} from a set of meters $\mathcal{P}\subseteq\mathcal{M}$, if and only if a unique estimate of $\mathcal{D}$ can be obtained from the measurements $\mathcal{P}$. That is, for two different vectors $\boldsymbol{\bar{\theta}}\neq \boldsymbol{\dot{\theta}}$, if
\begin{equation}
\mathbf{H}_{\{\mathcal{P}\},*} \cdot \boldsymbol{\bar{\theta}} = \mathbf{H}_{\{\mathcal{P}\},*} \cdot \boldsymbol{\dot{\theta}} = \mathbf{z}_{\mathcal{P}}
\end{equation}
holds for an arbitrary measurement vector $\mathbf{z}_{\mathcal{P}}$, then
\begin{equation}
\bar{\theta}_k = \dot{\theta}_k, \ \ \forall k\in \mathcal{D}.
\end{equation}
Likewise, a measured subnetwork $\bar{G}\left(\mathcal{P}\right)=\left(\mathcal{V}',\mathcal{E}'\right)$ is an \emph{observable subnetwork} if and only if all the unknown state variables $\mathcal{S}$ in the subnetwork is observable from $\mathcal{P}$, i.e.
\begin{equation}
\label{52}
rank\left(\mathbf{H}_{\left\{\mathcal{P}\right\},\left\{\mathcal{S}\right\}}\right)=|\mathcal{S}|,
\end{equation}
where $\mathcal{S}=\mathcal{V'}\setminus R$, with $R$ being the reference bus.

\textbf{Remark $1$:} It holds that $|\mathcal{P}|\geq |\mathcal{D}|$ if $\mathcal{D}$ is observable from $\mathcal{P}$. We refer to $\mathcal{P}$ as a \emph{basic measurement set} of $\mathcal{D}$, if $\mathcal{D}$ is observable from $\mathcal{P}$ and $|\mathcal{P}|=|\mathcal{D}|$. Notice that not all $\mathcal{D}$'s have a basic measurement set. From (\ref{52}), $\mathcal{P}$ contains at least a basic measurement set of $\mathcal{V}'$ when $\bar{G}\left(\mathcal{P}\right)$ is an observable subnetwork. Besides, $\bar{G}$ must include the reference bus $R$, i.e. $R\in \mathcal{V'}$, since otherwise $rank\left(\mathbf{H}_{\left\{\mathcal{P}\right\},\left\{\mathcal{S}\right\}}\right)<|\mathcal{S}|$. Note that the conventional definition of network observability is a special case with $\mathcal{D}=\mathcal{I}$ and $\mathcal{P}=\mathcal{M}$.

Now, we are ready to establish the equivalence between state observability and state estimate protection criterion.

\textbf{Theorem $1$:} Protecting a set of meter measurements $\mathcal{P}$ can defend a set of state variables $\mathcal{D}$ against undetectable attack, if and only if $\mathcal{D}$ is observable from $\mathcal{P}$.

\emph{Proof:} We first prove the \emph{if} part. When $\mathcal{D}$ is observable from $\mathcal{P}$, there must exist an observable subnetwork  $\bar{G}\left(\mathcal{\bar{P}}\right) = \left(\mathcal{\bar{V}},\mathcal{\bar{E}}\right)$ that includes $\mathcal{D}$, i.e. $\mathcal{D}\subseteq \mathcal{\bar{V}}$ and $\mathcal{\bar{P}} \subseteq \mathcal{P}$. From (\ref{52}), we have $rank\left(\mathbf{H}_{\left\{\mathcal{\bar{P}}\right\},\left\{\mathcal{\bar{S}}\right\}}\right)=|\mathcal{\bar{S}}|$, where $\mathcal{\bar{S}}=\mathcal{\bar{V}}\setminus R$. Then, the solution of $\mathbf{c}$ to $\mathbf{H}_{\left\{\mathcal{\bar{P}}\right\},*}\mathbf{c} = \mathbf{0}$ is $\mathbf{c}=\left(\mathbf{0},\mathbf{c}_{\mathcal{I}\setminus \mathcal{\bar{S}}}\right)^\intercal$, where $\mathbf{c}_{\mathcal{I}\setminus \mathcal{\bar{S}}}$ is an arbitrary vector. That is, no undetectable attack can be formulated to compromise $\mathcal{\bar{S}}$ if $\mathcal{\bar{P}}$ is well protected. Since  $\mathcal{D}\subseteq \mathcal{\bar{S}}$ and $\mathcal{\bar{P}}\subseteq \mathcal{P}$, this completes the proof of the \emph{if} part.

We then show the \emph{only if} part. That is, there exists an undetectable attack to compromise $\mathcal{D}$ if $\mathcal{D}$ is unobservable from $\mathcal{P}$. From Definition $2$, there exists a $\mathbf{z}_{\mathcal{P}}$ and two different state vectors $\boldsymbol{\bar{\theta}}$ and $\boldsymbol{\dot{\theta}}$, satisfying
\begin{equation}
\mathbf{z}_{\mathcal{P}} = \mathbf{H}_{\{\mathcal{P}\},*} \boldsymbol{\bar{\theta}} = \mathbf{H}_{\{\mathcal{P}\},*} \boldsymbol{\dot{\theta}}
\end{equation}
and $\bar{\theta}_k\neq \dot{\theta}_k$ for some $k\in \mathcal{D}$. By letting $\mathbf{c} = \boldsymbol{\bar{\theta}} - \boldsymbol{\dot{\theta}}$, we have $\mathbf{H}_{\{\mathcal{P}\},*} \mathbf{c}=\mathbf{0}$ and $c_k\neq 0$. In other words, an attacker can introduce non-trivial error $c_k$ to state variable $k\in \mathcal{D}$ without the need to compromise any protected meter in $\mathcal{P}$. Therefore, an undetectable attack $\mathbf{a}=\mathbf{H}\mathbf{c}$ can compromise state $\theta_k$ without being detected. $\hfill \blacksquare$

\textbf{Remark $2$:} Theorem $1$ indeed provides an equivalent condition as (\ref{11}) in protecting a set of state variables from the perspective of network observability. This will help to develop graphical algorithms in the following subsections. From Theorem $1$, we see that all the unknown state variables to be defended, i.e. $\mathcal{D}$, are included in an observable subnetwork constructed from a set of protected meters. In the following subsection, we find that the optimal observable subnetwork has an interesting Steiner tree structure.

\subsection{Graphical equivalence of optimal protection}
The power network observability analysis in \cite{1980:Krumpholz} showed a connection between network observability and a spanning tree structure. The idea is briefly covered in Proposition $1$.

\textbf{Proposition $1$:} The measured full network $\bar{G}\left(\mathcal{M}\right)=\left(\mathcal{V},\mathcal{E}\right)$ is observable if and only if the graph defined on $\bar{G}$ contains a spanning tree, where each edge of which is mapped to a meter according to the following rules,
\begin{enumerate}
  \item an edge is mapped to a flow meter placed on it, if any;
  \item an edge without a flow meter is mapped to an injection meter that measures it;
  \item different edges are mapped to different meters in $\mathcal{M}$.
\end{enumerate}

\emph{Proof:} See the proof in \cite{1980:Krumpholz}. $\hfill \blacksquare$

Proposition $1$ states that any basic measurement set of $\mathcal{V}$ can be mapped to a spanning tree in the measured full graph. On the other hand, a measured subnetwork $\bar{G}\left(\mathcal{P}\right)=\left(\mathcal{\bar{V}},\mathcal{\bar{E}}\right)$, where $\mathcal{P}\subseteq \mathcal{M}$, can also be considered as a closed network whose observability is only related to the components within $\bar{G}\left(\mathcal{P}\right)$. Therefore, there also exists a measurement-to-edge mapping in an observable subnetwork, specified as following.

\textbf{Corollary $1$:} A measured subnetwork $\bar{G}\left(\mathcal{P}\right)=\left(\mathcal{\bar{V}},\mathcal{\bar{E}}\right)$ is observable if and only if the graph defined on $\bar{G}\left(\mathcal{P}\right)$ contains a tree that connects all vertices in $\mathcal{\bar{V}}$, where each edge of the tree is one-to-one mapped to a unique meter in $\mathcal{P}$ that takes its measurement.

\emph{Proof:} The proof follows by replacing $\mathcal{M}$ with $\mathcal{P}$ in Proposition $1$. $\hfill \blacksquare$

From Remark $2$ and Corollary $1$, we see that the unknown state variables to be defended are indeed contained in a tree constructed from a protected meter measurement set. Therefore, we propose the following \emph{minimum measured Steiner tree} (MMST) problem in a graph that is equivalent to the optimal state protection problem (\ref{98}).

\begin{figure}
\centering
  \begin{center}
    \includegraphics[width=2.5in]{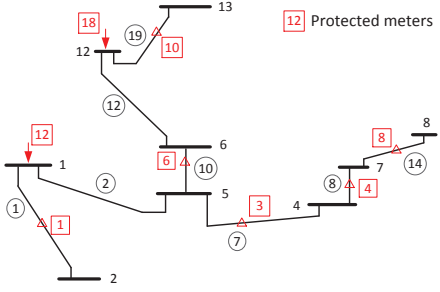}
  \end{center}
  \caption{An illustration of MMST from the IEEE $14$-bus testcase.}
  \label{73}
\end{figure}

\textbf{MMST problem:} Given the measured full graph $\bar{G}\left(\mathcal{M}\right)=\left(\mathcal{V},\mathcal{E}\right)$. To protect a set of state variables $\mathcal{D}$ with a minimum cost, the MMST problem finds a shortest Steiner tree $T^*=\left(\mathcal{V}^*,\mathcal{E}^*\right)$ (with the minimum number of edges) and a set of meters $\mathcal{P}^*\subseteq \mathcal{M}$ that satisfy the following conditions.
\begin{enumerate}
  \item $\mathcal{V}^*$ is the set of all vertices measured by $\mathcal{P}^*$;
  \item $\mathcal{D}\subset \mathcal{V}^*$ and $R\in \mathcal{V}^*$;
  \item each edge in $\mathcal{E}^*$ is one-to-one mapped to a unique meter in $\mathcal{P}^*$ that takes its measurement.
\end{enumerate}
Then, the set of meters $\mathcal{P}^*$ is the optimal solution to (\ref{98}).

We name the problem as a Steiner tree problem, instead of spanning tree, because  $T^*$ in general connects only a subset of vertices in the measured full graph. The three conditions ensure that all the unknown state variables in $T^*$, including $\mathcal{D}$, are observable from $\mathcal{P}^*$. We present an example from Fig. $\ref{65}$ to illustrate the structure of a MMST. We assume that $\mathcal{D}=\left\{v_8,v_{12}\right\}$ and $v_1$ is the reference bus. The optimal protected meters set $\mathcal{P}^*=\left\{r_1,r_3,r_4,r_6,r_8,r_{10},r_{12},r_{18}\right\}$ is obtained from exhaustive search. The corresponding minimum Steiner tree $T^*$ is plotted in Fig. $\ref{73}$. We see that conditions $1$) and $2$) are clearly satisfied. Condition $3$ is satisfied by mapping edges $e_{2}$ and $e_{12}$ to injection meters $r_{12}$ and $r_{18}$, and the other edges in $\mathcal{E}^*$ to the flow measurements placed on them.

We show that the MMST problem is \emph{NP-hard} by considering a special case where flow meters are installed at all edges of $\bar{G}\left(\mathcal{M}\right)=\left(\mathcal{V},\mathcal{E}\right)$. Then, any Steiner trees that include $R$ and $\mathcal{D}$ automatically satisfy the three conditions, i.e. by mapping each edge to the corresponding flow meter. In this case, the MMST problem becomes a standard minimum \emph{Steiner tree} (MST) problem, which finds the shortest subtree of the full graph that connects $R$ and all the vertices in $\mathcal{D}$. MST is a well-known \emph{NP-hard} problem. The time complexity of known exact algorithms increase exponentially with $|\mathcal{D}|$ or $|\mathcal{I}|-|\mathcal{D}|$ \cite{1992:Hwang}. Since MST is a special case of the MMST problem, the MMST problem is also \emph{NP-hard} following the reduction lemma for computational complexity analysis. A special case of the MMST problem with $\mathcal{D}= \mathcal{I}$ is solved in \cite{1980:Krumpholz} and \cite{1986:Barglela} with time complexity $O\left(|\mathcal{V}||\mathcal{E}|\right)$. The special case is easy because $\mathcal{V}^*=\mathcal{V}$ holds automatically when all the state estimates are to be protected. The general MMST problem is much harder due to the combinatorial nature of possible $\mathcal{V}^*$.

\section{Graphical Methods for Optimal Protection}
In this section, we first introduce two exact solution methods to solve the MMST problem, including the SVE method and an MILP formulation. Then, a tree pruning heuristic is proposed to obtain an approximate solution in polynomial time.

\subsection{Steiner vertex enumeration algorithm}
A vertex $v$ in the Steiner tree solution $T^*=\left(\mathcal{V}^*,\mathcal{E}^*\right)$ is a \emph{terminal} if $v\in \mathcal{D}\cup R$, or a \emph{Steiner vertex} otherwise. The Steiner vertex enumeration (SVE) method enumerates the possible Steiner vertices $\mathcal{V}_0$ until a minimum observable subnetwork, including $\mathcal{V}_0$ and the terminals, is found. Then, $\mathcal{P}^*$ can be obtained by removing redundant measurements in the subnetwork using Gauss-Jordan elimination. A pseudo-code of the SVE is presented in Algorithm $1$. The time complexity of SVE is $O\left(2^{|\mathcal{I}|-|\mathcal{D}|}\right)$, which is computational infeasible in large scale power networks, e.g. a $118$-bus system. Therefore, we mainly use SVE as the performance benchmark to evaluate the correctness of the algorithms proposed in the following subsections.

\begin{algorithm}
\small
 \SetAlgoLined
 \SetKwData{Left}{left}\SetKwData{This}{this}\SetKwData{Up}{up}
 \SetKwRepeat{doWhile}{do}{while}
 \SetKwFunction{Union}{Union}\SetKwFunction{FindCompress}{FindCompress}
 \SetKwInOut{Input}{input}\SetKwInOut{Output}{output}
 \Input{$\mathcal{I}, \mathcal{D}, \mathcal{M}$, $R$}
 \Output{Minimum protected measurements $\mathcal{P}^*$ to defend $\mathcal{D}$}
      \Repeat{$rank\left(\mathbf{H}_{\{\mathcal{\bar{P}}\},\{\mathcal{\bar{S}}\}}\right)=|\mathcal{\bar{S}}|$}{
      Enumerate a set of Steiner vertices $\mathcal{V}_0 \subseteq \left\{\mathcal{I}\setminus \mathcal{D}\right\}$, from size $|\mathcal{V}_0|=0$ to $|\mathcal{I}|- |\mathcal{D}|$. Let $\mathcal{\bar{S}}= \mathcal{D}\cup \mathcal{V}_0$;

      Find the meters $\mathcal{\bar{P}}$ that measure only the buses in $\mathcal{\bar{S}}\cup R$\;
}
 $\mathcal{P}^*=$ a basic measurement set of $\mathcal{\bar{S}}$\;
\caption{Steiner vertex enumeration algorithm}
\end{algorithm}

\subsection{Mixed integer linear programming formulation}
In this subsection, we propose an MILP formulation to solve the MMST problem, which has much lower complexity than SVE by exploiting the optimal solution structure. Consider a digraph $\overrightarrow{G}=\left(\mathcal{V},\mathcal{A}\right)$ constructed by replacing each edge in the measured full graph $\bar{G}\left(\mathcal{M}\right)=\left(\mathcal{V},\mathcal{E}\right)$ with two arcs in opposite directions. We set the reference bus as the root and allocate one unit of demand to each vertex in $\mathcal{D}$. Commodities are sent from the root to the vertices in $\mathcal{D}$ through some arcs. Then, the vertices in $\mathcal{D}$ are connected to $R$ via the used arcs if and only if all the demand is satisfied. When we require using the minimum number of arcs to deliver the commodity, the used arcs will form a directed tree, referred to as a \emph{Steiner arborescence}. Evidently, the solution to the MMST problem can be obtained if we solve the following \emph{minimum measured Steiner arborescence} (MMSA) problem and neglect the orientations of the arcs. Without causing confusions, we say an arc $(i,j)$ is measured by a meter if the edge $[i,j]$ in $\bar{G}\left(\mathcal{M}\right)$ is measured by the meter.

\textbf{MMSA problem:} Given a digraph $\overrightarrow{G}=\left(\mathcal{V},\mathcal{A}\right)$, find the shortest arborescence $\overrightarrow{T}^*= \left(\mathcal{V}^*,\mathcal{A}^*\right)$ and a set of meters $\mathcal{P}^*\subseteq \mathcal{M}$ that satisfy the following conditions
\begin{enumerate}
  \item $\mathcal{V}^*$ is the set of all vertices measured by $\mathcal{P}^*$;
  \item $\mathcal{D}\subset \mathcal{V}^*$ and $R\in \mathcal{V}^*$;
  \item each arc in $\mathcal{A}^*$ is one-to-one mapped to a unique meter in $\mathcal{P}^*$ that takes its measurement.
\end{enumerate}

From condition $1$), if an arc in $\overrightarrow{T}^*$ is mapped to an injection meter, all the vertices measured by the injection meter must also be included in the arborescence like the terminals, as if an extra demand is allocated at these vertices. To distinguish from the actual demand at $\mathcal{D}$, we refer to the extra demand induced by the use of injection meters as \emph{pseudo demand}. Then, the MMSA problem is to satisfy both the actual and pseudo demand using minimum number of arcs.

For an arc $(i,j)\in \mathcal{A}$, let $x_{ij}$ be a binary variable with $x_{ij}=1$ indicating that the arc is included in $\overrightarrow{T}^*$ and $0$ otherwise. $y_{ij}$ denotes the total amount of commodity through $(i,j)$. $z_{ij}$ be a binary variable with $z_{ij}=1$ indicating that the injection meter at vertex $i$ is mapped to arc $\left(i,j\right)$ or $(j,i)$, and $0$ otherwise. Then, an MILP formulation of the MMSA problem is

\begin{subequations}
\label{27}
\begin{align}
& \underset{\mathbf{X},\mathbf{Y},\mathbf{Z}}{\text{min}} & & \sum_{\left(i,j\right)\in \mathcal{A}}x_{ij} + \frac{1}{w}\sum_{\left(i,j\right)\in \mathcal{A}}z_{ij} \label{26}\\
&\text{s. t. }  & & x_{ij}\geq \frac{y_{ij}}{w},  \;\; \forall \left(i,j\right)\in \mathcal{A} \label{21}\\
&  & & \mathbf{1}_E(i,j) + z_{ij} + z_{ji}\geq x_{ij}, \; \forall \left(i,j\right)\in \mathcal{A} \label{22}\\
&  & & \sum_{\left(i,j\right)\in \mathcal{A}} z_{ij}\leq \mathbf{1}_V(i), \;\; \forall i\in \mathcal{V} \label{23}\\
&  & & \sum_{\left(i,j\right)\in \mathcal{A}}y_{ij}- \sum_{\left(j,k\right)\in \mathcal{A}}y_{jk}= d(j), \forall j\in \mathcal{V} \setminus R \label{24}\\
&  & & x_{ij},z_{ij}\in \left\{0,1\right\}, \ y_{ij}\geq 0, \forall (i,j)\in \mathcal{A}. \label{25}
\end{align}
\end{subequations}
Here, $w$ is chosen as a large positive number such that $w>\sum_{\left(i,j\right)\in A}z_{ij}$ and $w > y_{ij}$ always hold. $\mathbf{1}_E(i,j)$ and $\mathbf{1}_V(i)$ are two binary indicator functions, where $\mathbf{1}_E(i,j)=1$ if a flow meter is available at edge $[i,j]$ and $\mathbf{1}_V(i)=1$ if an injection meter is available at $v_i$. $d(j)$ is the demand at vertex $j$, where
\begin{equation*}
d(j)= \begin{cases}
1+\sum_{\left(j,k\right)\in \mathcal{A}} z_{jk} + \sum_{\left[k,j\right]\in \mathcal{E}}\sum_{\left(k,s\right)\in \mathcal{A}} z_{ks} &   j\in \mathcal{D}\\
\sum_{\left(j,k\right)\in \mathcal{A}} z_{jk} + \sum_{\left[k,j\right]\in \mathcal{E}}\sum_{\left(k,s\right)\in \mathcal{A}} z_{ks}   &   j\notin \mathcal{D}.\\
\end{cases}
\end{equation*}
For $j\notin \mathcal{D}$, $d(j)$ is the total pseudo demand. Otherwise, one extra unit of actual demand is counted as well.

As we can see, there are two terms in (\ref{26}), each corresponding to one objective. The first term is to minimize the total number of arcs included in the arborescence. The second term is to minimize the number of injection measurements. Notice that the first objective is primary, as the second term in (\ref{26}) is always dominated by the first one due to the scaling factor $1/w$, which makes the second term always less than $1$. As such, (\ref{26}) is to minimize the total number of arcs in the arborescence, and meanwhile eliminating redundant injection measurements, such as the case when two injection measurements are assigned to the same arc. Constraint $(\ref{21})$ forces arc $(i,j)$ to be included in $\overrightarrow{T}^*$ if any commodity flow passes through $(i,j)$. Constraint (\ref{22}) and (\ref{23}) ensure that each arc $(i,j)$ included in $\overrightarrow{T}^*$ has at least one measurement assigned to it and each injection measurement can only be assigned to at most one arc. The flow conservative constraint (\ref{24}), together with $(\ref{21})$, forces the selected arcs to form an arborescence rooted at the reference vertex and spanning all vertices with positive demand. Once the optimal solution to (\ref{27}) is obtained, we can restore the optimal solution $\mathcal{P}^*$ to the MMST problem by including:
\begin{enumerate}
  \item injection measurement on bus $i$ if $z_{ij}=1$, $\forall (i,j)\in \mathcal{A}$;
  \item flow measurement on arc $\left(i,j\right)$, if $x_{ij}=1$ and $z_{ij}=z_{ji}=0$, $\forall \left(i,j\right)\in \mathcal{A}$. That is, the arcs in $\overrightarrow{T}^*$ not mapped to any injection measurement.
\end{enumerate}

Extensive experiments in the simulation section show that the MILP formulation always obtains the same optimal solution as the SVE algorithm. Besides, the MILP significantly reduces the computational complexity by exploiting the solution structure. For instance, a problem in a $57$-bus system that is computationally infeasible by the SVE algorithm can now be solved by the MILP within minutes. Nonetheless, the computational complexity of the state-of-art MILP algorithms, such as branch and bound and cutting-plane method, etc, still grows exponentially with the problem size. We observe from simulations that it takes excessively long time to solve the problem in a $300$-bus power system.

\subsection{Tree pruning heuristic}
To tackle the intractability of the problem, we propose a tree-pruning based heuristic (TPH) that finds an approximate solution in polynomial time. We refer to a tree $T=\left(\mathcal{\bar{V}},\mathcal{\bar{E}}\right)$, along with a set of measurement $\mathcal{\bar{P}}$, a \emph{feasible measured tree} if  $T$ and $\mathcal{\bar{P}}$ satisfy the conditions of the MMST problem. Our observation is that, although it is hard to find a MMST, it is relatively ``easy" to find a feasible tree that includes all the vertices in the graph using the techniques in \cite{1980:Krumpholz}. Starting from a feasible measured tree that spans all vertices in the measured full graph, our TPH method iteratively prunes away redundant vertices and updates the feasible tree, until a shortest possible tree is obtained. A pseudo-code is provided in Algorithm $2$. The TPH consists of multiple rounds of pruning operations. Here, we explain one round of pruning, which corresponds to line $2$-$8$ in the pseudo-code, in the following $4$ steps.

\begin{algorithm}
\small
 \SetAlgoLined
 \SetKwData{Left}{left}\SetKwData{This}{this}\SetKwData{Up}{up}
 \SetKwRepeat{doWhile}{do}{while}
 \SetKwFunction{Union}{Union}\SetKwFunction{FindCompress}{FindCompress}
 \SetKwInOut{Input}{input}\SetKwInOut{Output}{output}
 \Input{$\bar{G}\left(\mathcal{M}\right)=\left(\mathcal{V},\mathcal{E}\right)$, $\mathcal{D}$, $R$, $K$ }
 \Output{Minimum protected measurements $\mathcal{P}^*$ to defend $\mathcal{D}$}
 \textbf{initialization:}  $\mathcal{\bar{V}}=\mathcal{V}$\;
      \Repeat{$W=|T^*|$}{
      Let $W=|\mathcal{\bar{V}}|$. Find $K$ basic measurement sets of $\mathcal{\bar{V}}$, denoted by $\mathcal{\bar{P}}^k$, $k=1,..,K$. For each $\mathcal{\bar{P}}^k$, construct a feasible measured trees $T_k$\;
      \For{\emph{each} $T_k$}{
            Starting from $R$ to all leaf vertices, find the largest prunable subset $C_s^*(i)$ for each $v_i$. Update $T_k=$ $T_k\setminus\left\{C^*_s(i)\cup D(C^*_s(i))\right\}$ until each vertex in $T_k$ is either processed or pruned\;
      }
      Select the minimum trees $T^*$ and update $\mathcal{\bar{V}}$ by letting $\mathcal{\bar{V}}=$ the vertices in $T^*$\;
}
 $\mathcal{P}^*=$ the remaining measurements corresponding to $T^*$\;
 \caption{Tree pruning heuristic algorithm}
\end{algorithm}

\textit{Step 1: Feasible tree generation.} For a set of vertices $\mathcal{\bar{V}}$ (initially set to be $\mathcal{V}$), we generate $K$ feasible edge-measured trees that span all the vertices in $\mathcal{\bar{V}}$, where $K$ is a tunable parameter (lines $3$-$4$). In this step, we first find the meters that measure only the vertices in $\mathcal{\bar{V}}$. This can be easily performed by examining in $\mathbf{H}$ whether all the non-zero elements in a row lie in the columns corresponding to the state variable set $\mathcal{\bar{V}}\setminus R$. For instance, for $\mathcal{\bar{V}}=\left\{v_1,v_2,v_4,v_5\right\}$ and $R=v_1$ in Fig. \ref{61}, the selected meters are $\left\{r_1,r_2,r_4,r_6\right\}$. Among the selected meters, we find $K$ basic measurement sets of $\mathcal{\bar{V}}\setminus R$, denoted by $\mathcal{\bar{P}}^k$ ($k=1,..,K$), using Gauss-Jordan elimination. Then, we construct $K$ feasible spanning trees, one for each $\mathcal{\bar{P}}^k$, using the max-flow method given in the Appendix. The $K$ feasible spanning trees are denoted by $T_k=\left(\mathcal{\bar{V}},\mathcal{\bar{E}}^k\right)$, $k=1,..,K$.

\begin{figure}
\centering
  \begin{center}
    \includegraphics[width=3in]{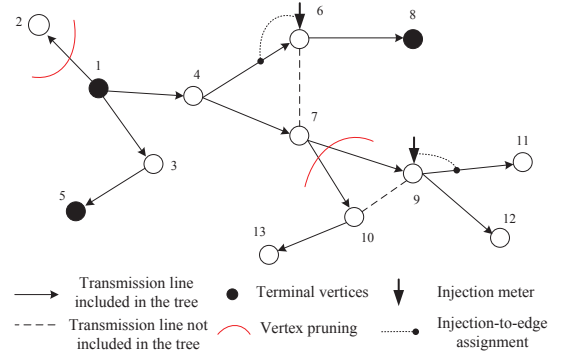}
  \end{center}
  \caption{A measured feasible tree. $\left\{v_1,v_5,v_8\right\}$ are the terminals and $v_1$ is the reference. Two marked edges ($[4,6]$ and $[9,11]$) are mapped to injection meters and the other unmarked edges are mapped to flow meters.}
  \label{63}
\end{figure}

\textit{Step 2: Vertex identification.} For each tree $T_k$, we identify the child and descendant vertices of each vertex (included in line $5$-$6$ in Algorithm $2$). This can be achieved by constructing a directed tree from the root to all leaf vertices. If there is an arc $(i,j)$, we say $v_j$ is a child of $v_i$, denoted by $v_j\in C\left(i\right)$. In general, if there exists a path from $v_i$ to $v_j$, we refer to $v_j$ as a descendent of $v_i$, denoted by $v_j\in D(i)$. In Fig. $\ref{63}$, for instance, $v_6$ and $v_7$ are the child vertices of $v_4$, while $v_6$ to $v_{13}$ are all descendent vertices of $v_4$. In practice, the descendent vertex identification can be achieved using breadth-first-search starting from the root.

\textit{Step 3: Tree pruning.} For each $T_k$, we start from the root to the leaf vertices to prune away redundant vertices (line $5$-$6$ in Algorithm $2$). For a vertex $v_i$, we find the largest prunable subset $C_s^*(i)\subseteq C(i)$, such that the residual tree is still a feasible measured tree after all the vertices in $\{C_s(i)\cup D(C_s(i))\}$ are pruned. In particular, $\{C_s(i)\cup D(C_s(i))\}$ can be pruned if:
\begin{enumerate}
  \item $\{C_s(i)\cup D(C_s(i))\}$ contains no terminal vertex,
  \item the deletion of $\{C_s(i)\cup D(C_s(i))\}$ will remove all the edges mapped to injection meters that measure any vertex in $\{C_s(i)\cup D(C_s(i))\}$.
\end{enumerate}
This is because the first condition ensures the all the state variables to be protected is still included in the tree. The second condition guarantees that the vertices in the residual tree are only measured by the remaining measurements. The two conditions ensure that the residual tree is feasible to the MMST problem. Then, we update $T_k$ by removing all the vertices in $\{C_s^*(i)\cup D(C_s^*(i))\}$ and proceed to another vertex until each vertex in $\mathcal{\bar{V}}$ is either checked or pruned.

\textit{Step 4: Vertex update.} Let $|T_k|$ be the number of remaining vertices in $T_k$. Then, we select among the $K$ trees the one with minimum vertices, denoted by $T^*$. If $|T^*|=|\mathcal{\bar{V}}|$, i.e. no vertex is removed for all the $K$ trees, we terminate the algorithm and output $\mathcal{P}^*$ as the remaining meters in $T^*$ (line $7$-$9$). Otherwise, we first update $\mathcal{\bar{V}}$ as the remaining vertices in $T^*$ and start another round of pruning from Step $1$).

In Fig. $\ref{63}$, we present an example to illustrate the TPH, where a feasible tree contains $12$ vertices is presented. Starting from the root $v_1$, among the three child vertices of $v_1$, only $v_2$ can be pruned, since the descendent vertices of either $v_3$ or $v_4$ contain terminal vertex. After pruning $v_2$, we proceed to check $v_3$, whose only child vertex $v_5$ is a terminal. Then, we check $v_4$, where neither of its child vertices $v_6$ and $v_7$ can be pruned separately or together. On one hand, this is because $v_6$ contains terminal as its descendent vertices. On the other hand, the removal of $v_7$ does not remove the edge $\left[4,6\right]$, which is mapped to the injection meter at $v_6$ that measures $v_7$. For $v_7$, however, all of its descendent vertices can be pruned following the two pruning conditions. Up to now, we have finished the first round of pruning. Then, we use the remaining vertices $\left\{v_1,v_3,v_4,v_5,v_6,v_7,v_8\right\}$ to generate new feasible trees, if any, and repeat the pruning operations iteratively until no vertex can be further pruned.

The purpose of introducing the parameter $K$ is because the final output $\mathcal{P}^*$ is closely related to the tree's topology obtained in Step $1$. Intuitively, with larger $K$, we have a larger chance to obtain a smaller $|\mathcal{P}^*|$ but also consume more computations. The proper choice of $K$ will be discussed in Simulations. The correctness of TPH is obvious from the following facts: $1$) the $K$ residual trees are always feasible measured tree; $2$) the size of the minimum residual tree is non-increasing during the iterations; $3$) $|\mathcal{P}^*|$ equals the size of the minimum residual tree. There are at most $|\mathcal{I}|-|\mathcal{D}|$ rounds of pruning. In each round, $K$ trees are pruned and each takes $O\left(|\mathcal{I}|^3\right)$ time complexity, dominated by the Gauss-Jordan elimination computation. The overall time complexity is $O\left(K|\mathcal{I}|^4\right)$, which is considered efficient even for very large scale power systems.

\section{Discussions of Application Environments}
In this section, we discuss the possibility of extending the proposed algorithms to some interesting application scenarios. The topics we consider include: the integration of phasor measurement units (PMUs) into state estimation, the applicability to AC state estimation model and the extension to achieve incremental state variable protection. Interestingly, we find that our proposed algorithms can fit in all the considered scenarios with minor modifications.

\subsection{Integration with phasor measurement unit}
Recently, the introduction of more sophisticated measurement components has largely improved the accuracy and reliability of state estimation. One such device is the phasor measurement unit (PMU). Combined with GPS technology, PMUs can provide direct real-time voltage phasor measurement, i.e. voltage amplitude and phase angle,\footnote{There also exists other type of PMUs that can provide current phasors of all the incident branches besides bus voltage phasors. We do not include them into consideration in this paper because they are inconsistent with our notion of a ``measurement", which provides only one reading at a time. However, we consider this problem as a future work.} with high precision and short measurement periodic time \cite{1996:Zivanovic}. In other words, any bus with a PMU installed does not need to estimate its voltage phasor if the device has a credible precision. There have been a number of studies on the PMU deployment to improve power network observability \cite{2008:Chakrabarti,2013:Giani}. However, although the introduction of PMUs can be dated back to the $1980$s, its deployment had been in a slow pace until the past decade when a series of severe blackout experienced all around the world \cite{2011:Exposito}. Nowadays, the available PMUs alone are still not sufficient to guarantee the observability of entire power network. In practice, we need to rely on the mixed measurements provided by both PMUs and the conventional SCADA system to derive the state estimates \cite{2009:Abbasy}.

Interestingly, our proposed algorithms can be easily extended to protect state estimation when PMUs are used. Note that the state variable of a tagged bus cannot be compromised by attacks if a secured PMU is installed at the bus.\footnote{PMU is normally required to be installed at the reference bus to avoid the confusions due to the absolute voltage phasor measurements.} This is equivalent to installing a secured flow meter between the tagged bus and the reference bus. If there exists no such power line connecting the two buses, a pseudo transmission line can be added to facilitate the calculation of the MMST problem. Then, the proposed protection algorithms can be directly applied to solve the MMST problem. The only modification needed is that injection meters cannot be mapped to a dashed edge in the Steiner tree solution, because they do not measure the dashed edges in real system. The modification can be easily made in the constraints on $z_{ij}$ in the MILP formulation (\ref{27}) by defining $z_{ij}=0$ if a dashed edge $e_{ij}$ is made up by a PMU. For the TPH, the pruning rules need slight modification due to the change of mapping rule of injection meters in the presence of PMUs. The details are omitted here to avoid the repetition of presentations.

\begin{figure}
\centering
  \begin{center}
    \includegraphics[width=0.4\textwidth]{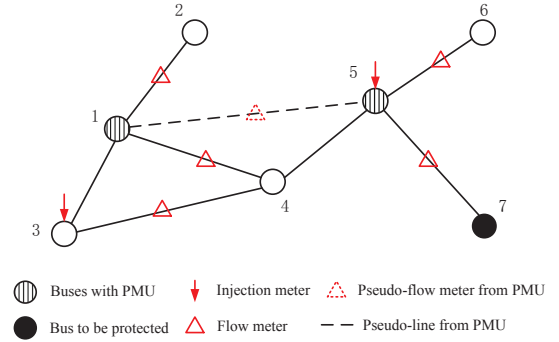}
  \end{center}
  \caption{Integration of PMUs in state estimation protection.}
  \label{pmu}
\end{figure}

We provide an illustrative example in Fig. $\ref{pmu}$, where a graph is extracted from a $7$-bus power network. Bus $1$ is the reference bus and PMUs are available at bus $1$ and $5$. The solid edges are the actual transmission lines in the power network. The dashed edge connecting bus $1$ and $5$ is made up by the PMU at bus $5$, where a pseudo-flow meter of random direction is placed on edge $e_{15}$. As discussed above, in any Steiner tree solution, the injection meter at bus $5$ cannot be mapped to the dashed edge $e_{15}$ made up by the PMU. Since now we have formulated an equivalent problem with only power flows/injections as the measurements, the proposed tree construction algorithms in Section IV can be directly applied. Suppose that the state variable of bus $7$ is to be protected, a Steiner tree can be constructed by edges $\left\{e_{15},e_{57}\right\}$, which are mapped to the pseudo-flow meter on edge $e_{15}$ (from the PMU at bus $5$) and the flow meter on edge $e_{57}$, respectively. Then, bus $7$ can be defended if the PMU at bus $5$ and the flow meter on $e_{57}$ are protected.

\subsection{Application to AC state estimation protection}
Unlike the linear DC power flow model, the measurement functions of AC power flow model consist of non-linear and coupled active and reactive power flow measurements. Meanwhile, the voltage amplitudes are also considered as the state variables in AC power flow model. Specifically, the active and reactive power flows on a power line connecting bus $i$ and $j$ are functions are
\begin{equation}
\label{AC}
  \begin{aligned}
  P_{ij} &= V_i^2 \cdot g_{ij} - V_iV_j\left[ g_{ij} \cos(\theta_i-\theta_j) + b_{ij} \sin(\theta_i-\theta_j)\right]\\
  Q_{ij} &= -V_i^2 \cdot b_{ij} + V_iV_j \left[ b_{ij} \cos(\theta_i-\theta_j) - g_{ij} \sin(\theta_i-\theta_j)\right],
  \end{aligned}
\end{equation}
where $V_i$ is the voltage amplitude at bus $i$, $g_{ij}$ and $b_{ij}$ are the conductance and susceptance of the power line (neglecting the shunt elements). Besides, the injection measurements at a bus are merely the sum of power flows in the incident branches. The AC state estimation is commonly performed in an iterative manner using the Newton's method \cite{2004:Abur}. False-data injection attack to AC state estimation is much harder than to the DC counter part. On one hand, both the active and reactive flows measurements need to be compromised. On the other hand, the attacker also needs to know the estimated values of state variables to calculate the attack parameter \cite{2012:Hug}. This basically requires the knowledge of all the real-time measurement readings.

Despite the apparent differences, we find that the proposed algorithms can still be applied to protect AC state estimation if the attackers only compromise voltage phase angle variables as in the DC model. In particular, the proposed methods remain both valid and optimal (for exact algorithms) in protecting state variables in AC state estimation. From the attackers' perspective, given $V_i's$ are constant (assumed untouched by attackers), we notice that the power flow measurements in (\ref{AC}) are only determined by phase angles differences, which is the same as in DC power flow model. For example, suppose that an attacker wants to perform an undetectable attack to compromise the phase angle variable of bus $5$ in Fig. \ref{pmu} (we assume the meters measure both active and reactive flows/injections and the PMUs are removed), by introducing the same error $c$ to the phase angle variables of bus $4$ to $7$. From (\ref{AC}), the attacker does not need to compromise the flow meters on $e_{56}$ and $e_{57}$, and the injection meter at bus $5$. However, it is necessary for the attacker to compromise the readings of the boundary meters, i.e. the flow meters on edge $e_{14}$ and $e_{34}$, and the injection meter on bus $3$. The compromised measurements will result in a biased estimate produced by the system operator's AC state estimator when the system is observable, i.e. the AC state estimation converges to a unique solution under any set of consistent measurements. In general, the attacker needs to find a cut that separates a tagged bus $k$ and the reference bus, introducing the same error $c$ to the subgraph that contains the bus $k$ and zero error to buses in the other subgraph that contains the reference bus. Then, the attacker only needs to compromise the meters, either flow or injection meters, which measure the buses on boundary.

Conversely, if a minimum measured Steiner tree is constructed by edges mapped to secured meter measurements from the reference to bus $k$, no undetectable attack can be performed. This is because any attack formulation by cut will require the attacker to compromise at least one secured meter measurement. Therefore, our proposed method for DC state estimation model remains valid and optimal in AC state estimation model. However, if attackers also compromise voltage amplitude state variables, our methods may still be valid but no longer optimal. This is because the readings of flow meters are now determined by the absolute values, rather than the difference of voltage amplitudes. More detailed analysis in AC state estimation protection will be considered as a future working direction.

\subsection{Extension to incremental state variable protection}
Another interesting extension of the proposed algorithms is to achieve incremental protection. Eventually, the system operator may want to protect all the state variables in the power system. However, due to the temporary limited budget and lengthy security installation time in a large scale power network, we may only be able to install security devices on a set of meters to protect a subset of state variables first. Later, we can extend the coverage to protect the other state variables given the already protected meters, until all state variables are protected. In fact, our proposed algorithms can be extended to achieve such incremental protection. The intuitive idea is to ``grow" a new feasible tree on top of the existing feasible tree to reaches more vertices to be protected.

Suppose that a set of state variables $\mathcal{D}_1$ has been defended by protecting a set of meters $\mathcal{P}_1$. A feasible tree $T_1$ can therefore be constructed using the maximum-flow technique introduced in the Appendix. By doing so, we also obtain the mapping between the measurement and edges. Assume that we want to extend the coverage to defend another set of state variables $\mathcal{D}_2$, i.e. $\mathcal{D}_1 \bigcap \mathcal{D}_2 =\emptyset$, given the protected meters $\mathcal{P}_1$. Notice that the choice of $\mathcal{D}_1,\mathcal{D}_2,...$ can be made arbitrary by the system operator. Intuitively, we need to find minimum number of edges, as well as the mapped meter measurements, to connect the vertices in $\mathcal{D}_2$ to the current feasible tree $T_1$. For the MILP formulation in (\ref{27}), we can first add to the constraints $x_{ij}=1$ and $z_{ij}=1$ for those edges and injection meter in the existing feasible tree $T_1$. That is, $x_{ij}=1$ if edge $e_{ij}$ is included in $T_1$; $z_{ij}=1$ if the injection meter at bus $v_i$ is mapped to the edge $e_{ij}$. Then, a new minimum measured Steiner tree (MMST) as well as the new meter set $\mathcal{P}_2$ to be protected can be calculated using the optimization in (\ref{27}). This can be achieved by a simple replacement of $\mathcal{D}$ with $\mathcal{D}_2$, i.e. deliver one unit of demand to each vertex in $\mathcal{D}_2$. Similar calculations can be performed to defend $\mathcal{D}_3$, $\mathcal{D}_4$, $\cdots$, until all the state variables are protected. For the TPH, we merely need to add several new policies to make sure that the MMST generated in the previous iteration to defend the variable set $\mathcal{D}_{i-1}$ remain intact in the current iteration to defend another variable set $\mathcal{D}_{i}$. The detailed pruning policies are omitted here due to the scope of this paper. Notice that the number of meters needed to protect all the state variables equals to the number of state variables, i.e. the size of a basic measurement set as introduced in \cite{2010:Bobba}, since we always keep a feasible tree whose edge is one-to-one mapped to a secured meter measurement.

Before leaving this session, we want to emphasize that all the proposed algorithms can be built on top of the existing state estimation application in EMS/SCADA. This is because the proposed algorithms merely find out a minimum set of meter measurements to be protected without altering the algorithm of state estimation or BDD. Besides, the calculation of the proposed algorithms can be done offline, independent of real-time measurements.

\section{Simulation Results}
In this section, we use simulations to evaluate the proposed defending mechanisms. All the computations are solved in MATLAB on a computer with an Intel Core2 Duo $3.00$-GHz CPU and $4$ GB of memory. In particular, MatlabBGL package is used to solve some of the graphical problems\cite{2006:Gleich}, such as maximum-flow calculation, etc. Besides, Gurobi is used to solve MILP problems \cite{2013:Gurobi}. The power systems we considered are IEEE $14$-bus, $57$-bus and $118$-bus testcases, whose topologies are obtained from MATPOWER \cite{2007:Zimmerman} and summarized in Table \ref{stat}. All the systems are observable with the respective measurement placement. For illustration purpose, a measurements placement of the 14-bus system is plotted in Fig. $\ref{65}$. The measurement placements for 57-bus and 118-bus systems are omitted for the simplicity of expositions.

\begin{table}
\caption{Statistics of Different Power System Testcases}
\footnotesize
\begin{center}
\begin{tabular}{|c|c|c|c|}
 \hline
  No. of buses &   $14$-bus    & $57$-bus & $118$-bus \\ \hline
  No. of lines &   $20$        & $80$     & $186$  \\ \hline
  Total no. of measurements &   $19$  & $80$     & $180$ \\ \hline
  No. of inject measurements &   $8$  & $30$     & $70$ \\ \hline
  No. of flow measurements &   $11$   & $50$     & $110$ \\ \hline
  No. of unmeasured lines &   $2$   & $2$     & $7$ \\ \hline
\end{tabular}
\end{center}
\label{stat}
\end{table}

We first verify the correctness of the MILP formulation to solve the optimal state variable protection problem. This is achieved by comparing the solutions of MILP against those of SVE algorithm in a $14$-bus system. The reason we use the $14$-bus testcase is because the SVE algorithm becomes computational infeasible in a larger power network, such as $57$-bus testcase. Besides the measurement placement in Fig. $\ref{65}$, two other measurement placements in the $14$-bus testcase are used as well, given that the power network is observable from all the measurement placements. We select $k$ of the $13$ unknown state variables (bus $1$ being the reference bus) as $\mathcal{D}$ to test each measurement placement, where $k = \left\{1,2,4,7,10\right\}$. For each $k$, $20$ randomly selected $\mathcal{D}$'s are tested using MILP formulation. Each entry in Table \ref{MILPVALID} is the percentage (hit ratio) that the MILP formulation yields the same number of meters as the optimal solution obtained by the SVE algorithm. We see that MILP formulation obtains the optimal solution for all the experiments. This, together with extensive other simulations, verifies the correctness of MILP formulation in solving the optimal protection problem.

\begin{table}
\caption{Hit ratio of MILP formulation in $14$-bus testcase}
\begin{center}
\begin{tabular}{|c|c|c|c|c|c|}
 \hline
  $|\mathcal{D}|$        &   $1$        & $2$        &   $4$      &  $7$      &  $10$     \\ \hline
  Measurement set $1$    &   $100\%$    & $100\%$    &   $100\%$  &  $100\%$  &  $100\%$   \\ \hline
  Measurement set $2$    &   $100\%$    & $100\%$    &   $100\%$  &  $100\%$  &  $100\%$   \\ \hline
  Measurement set $3$    &   $100\%$    & $100\%$    &   $100\%$  &  $100\%$  &  $100\%$   \\ \hline
\end{tabular}
\end{center}
\label{MILPVALID}
\end{table}

\begin{figure}
\centering
  \begin{center}
    \includegraphics[width=3.2in]{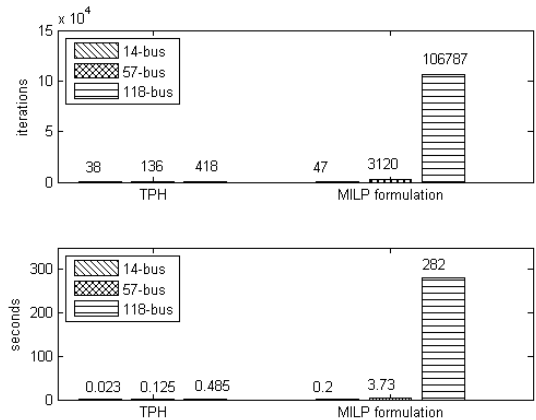}
  \end{center}
  \caption{Comparison of computational complexity for MILP and TPH. (a) The figure above shows the average number of iterations to obtain a solution; (b) the figure below shows the average CPU time to obtain a solution. }
  \label{66}
\end{figure}

We then evaluate the computational complexity of TPH in Fig. $\ref{66}$, where MILP is the benchmark for comparison. For TPH, we set the parameter $K=1$ and record the total number of vertices that are checked to produce a solution. For MILP, we record the number of nodes explored in the search tree by the branch-and-bound algorithm. Both numbers are the iterations consumed by the two methods to obtain a solution. Besides, we also record the CPU time for both methods. The results in Fig. $\ref{66}$ are the average performance of $50$ independent experiments. Without loss of generality, we randomly generate a $\mathcal{D}$ with size $|\mathcal{D}|=4$ in each experiment. In Fig. $\ref{66}$a, we show the average number of iterations for $14$-bus, $57$-bus and $118$-bus systems, respectively. We find that the iteration numbers are very close for both methods in the $14$-bus system, where TPH consumes $38$ iterations and the MILP consumes $47$ iterations to obtain a solution. However, the difference becomes more and more significant as the network size increases. The number of iterations of TPH increases by $11$ times as the network size increases from $14$ to $118$ buses. In vivid contrast, the iteration number of MILP increases rapidly by $2272$ times, from merely $47$ to $106787$. Similar results are also observed for the CPU time, where TPH takes only $0.485$ second to obtain a solution in $118$-bus system, while MILP consumes around $5$ minutes, which is $1410$ times slower than in the $14$-bus system. The booming computational complexity of the MILP method is due to the NP-harness of solving an MILP. It is foreseeable that the computational complexity of the MILP method will become extremely expensive as we further increase the network size. For instance, the projected CPU time of MILP to solve a problem in $300$-bus system is more than $5$ days, while it takes TPH less than $2$ seconds.

When protecting all the state variables, the state estimation protection problem in \cite{2010:Bobba} is a special case of ours. In this case, the proposed SVE and TPH algorithms indeed use the same Gauss-Jordan elimination technique proposed in \cite{2010:Bobba}, thus are of the same complexity. For the proposed MILP formulation, however, the complexity could be much higher due to the NP-harness of solving integer programming problems. Therefore, we do not recommend to using MILP to solve the special case that all the state variables are to be protected. Another point to mention is the impact of the redundancy in measurements. On one hand, the complexity of the MILP increases with the measurement redundancy. This is mainly because the number of variables $z_{ij}$ in the optimization problem (\ref{27}) will increase as the number of injection meter increases. On the other hand, the proposed TPH is not sensitive to measurement redundancy in the network. This is because its complexity is $O(K|I|^4)$, independent of the number of measurements.
\begin{table}
\caption{Performance of TPH and MILP in $57$-bus testcase}
\begin{center}
\begin{tabular}{|c|c|c|c|c|c|c|c|}
 \hline
  $|\mathcal{D}|$        &        $1$      & $4$            & $9$          &   $19$       &  $29$        &  $39$       & $49$    \\ \hline
  $|\mathcal{P}|$, $K=1$                  &   $11.8$        & $22.2$         & $30.3$       &   $39.5$     &  $46.3$      &  $51.6$     & $55.8$\\ \hline
  $|\mathcal{P}|$, $K=3$                  &   $10.7$        & $20.8$         & $28.0$       &   $37.0$     &  $43.0$      &  $48.8$     & $54.1$\\ \hline
  $|\mathcal{P}|$, $K=5$                  &   $9.9$         & $20.4$         & $27.8$       &   $36.7$     &  $42.5$      &  $47.9$     & $53.7$\\ \hline
  $|\mathcal{P}|$, $K=10$                 &   $9.7$         & $20.2$         & $27.3$       &   $36.3$     &  $42.1$      &  $47.6$     & $53.1$\\ \hline
  $|\mathcal{P}|$, $K=15$                 &   $9.4$         & $20.0$         & $26.8$       &   $35.9$     &  $41.7$      &  $47.3$     & $52.8$\\ \hline
  MILP ($|\mathcal{P}^*|$)      &   $8.8$         & $18.2$         & $25.4$       &   $34.6$     &  $40.7$      &  $46.2$     & $51.8$\\ \hline
  Gap                    &   $0.6$         & $1.8$         & $1.4$       &   $1.3$     &  $1.0$      &  $1.1$     & $1.0$\\ \hline
\end{tabular}
\end{center}
\label{TPHTABLE}
\end{table}

\begin{figure}
\centering
  \begin{center}
    \includegraphics[width=3.2in]{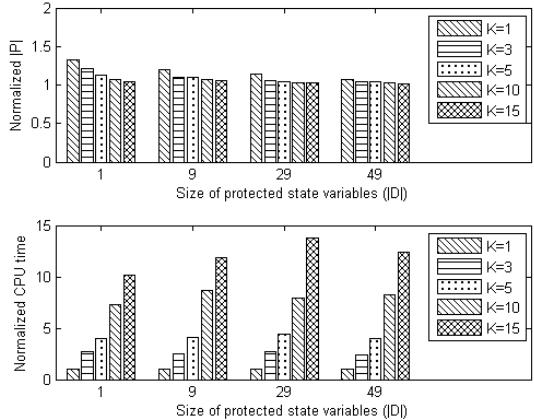}
  \end{center}
  \caption{Effect of $K$ to the performance of TPH in the $57$-bus system. (a) The figure above shows the solution size of TPH normalized by the optimal solution size obtained by MILP; (b) the figure below shows the CPU time of TPH normalized by the CPU time when $K=1$.}
  \label{67}
\end{figure}

We also investigate the impact of the parameter $K$ to the performance of TPH. By varying the values of $K$ and $|\mathcal{D}|$, we show in Table \ref{TPHTABLE} the average solution size $|\mathcal{P}|$ of TPH and MILP. Each entry of the table is the average performance of $50$ independent experiments. From the $2$nd to the $6$th rows, we see that better solution, i.e. smaller $|\mathcal{P}|$, is obtained with larger $K$. Compared with the optimal solution $\mathcal{P}^*$ obtained by MILP, TPH protects on average only $1.13$ more meters when $K=15$. The optimality gap is less than $10\%$ for all the cases. For better visualization, we plot the ratio $|\mathcal{P}|/|\mathcal{P}^*|$ for some selected $|\mathcal{D}|$'s in Fig. $\ref{67}a$. We notice that the ratio improves notably for small $|\mathcal{D}|$ as $K$ increases from $1$ to $15$. For instance, the ratio improves from $1.32$ to $1.04$ for $|\mathcal{D}|=1$. The improvement is especially notable when we change $K=1$ to $3$. However, the improvement becomes marginal as we further increase $K$, such as the case with $|\mathcal{D}|=49$, where the ratio only improves by $0.03$ from $K=1$ to $15$. We also plot in Fig. $\ref{67}b$ the CPU time normalized against the time consumed when $K=1$. We observe that the CPU time increases almost linearly with $K$, which matches our analysis in Section IV. Results in Fig. $\ref{67}$ indicate that we should select a proper $K$ to achieve a balance between the quality of approximate solution and computational complexity. In particular, a large $K$, such as $K=10$, should be used when $|\mathcal{D}|$ is small relative to $n$, i.e. $|\mathcal{D}|<0.1n$. Otherwise, a small $K$, such as $K=3$, should be used when $|\mathcal{D}|$ is relatively large.

\section{Conclusions}
In this paper, we used graphical methods to study defending mechanisms that protect a set of state variables from false-data injection attacks. By characterizing the optimal protection problem into a variant Steiner tree problem, we proposed both exact and approximate algorithms to select the minimum number measurements for system protection. The advantageous performance of the proposed defending mechanisms has been evaluated in IEEE standard power system testcases.


\appendix[Maximum-flow method for tree construction]

\begin{figure}
\centering
  \begin{center}
    \includegraphics[width=3in]{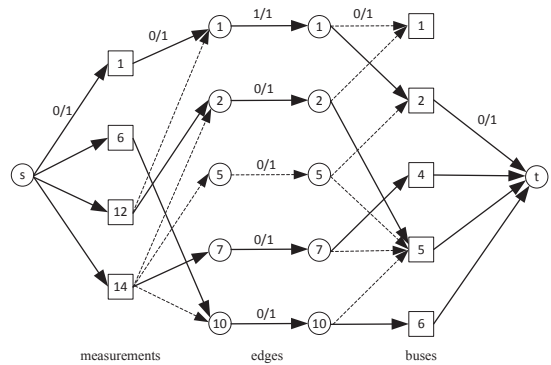}
  \end{center}
  \caption{Maximum-flow method for measurement tree construction. The solid arcs are the saturated arcs obtained in the final solution.}
  \label{55}
\end{figure}

We use an example in Fig. $\ref{65}$ to illustrate the method to obtain a feasible spanning tree. We consider a basic measurement set $\mathcal{\bar{P}}=\left\{r_1,r_6,r_{12},r_{14}\right\}$ of $\mathcal{\bar{V}}\setminus R$, where $\mathcal{\bar{V}}=\left\{v_1,v_2,v_4,v_5,v_6\right\}$ and $R=v_1$. The set of edges measured by $\mathcal{\bar{P}}$ is $\mathcal{\bar{E}}=\left\{e_1,e_2,e_5,e_7,e_{10}\right\}$. Then, a directed graph is constructed in Fig. $\ref{55}$, where $v_1$ is chosen as the root to construct the spanning tree. We select in advance an edge connected to the root, say $e_1$, in the final tree solution. This is achieved by setting both the lower and upper capacity bounds of the edge to be $1$. The other edges' lower and upper capacity bounds are set to be $0$ and $1$, respectively. Then, a maximum flow is calculated from $s$ to $t$. If the problem is feasible, i.e. the flow solution is $1$ in edge $e_1$, we obtain a measurement-to-edge mapping by observing the saturating flows in the graph. Otherwise, we select another edge connected to the root and recalculate the maximum flow problem. Since $\mathcal{\bar{V}}$ is observable from $\mathcal{\bar{P}}$, there is always a solution. In the above example, the final measurement-to-edge mapping is $\left\{r_1,r_6,r_{12},r_{14}\right\}\leftrightarrow \left\{e_1,e_{10},e_2,e_7\right\}$. Then, the edges obtained by the maximum flow calculation will form a tree that spans all vertices in $\bar{V}$.

\begin{IEEEbiography}
[{\includegraphics[width=1in,height=1.25in,clip,keepaspectratio]{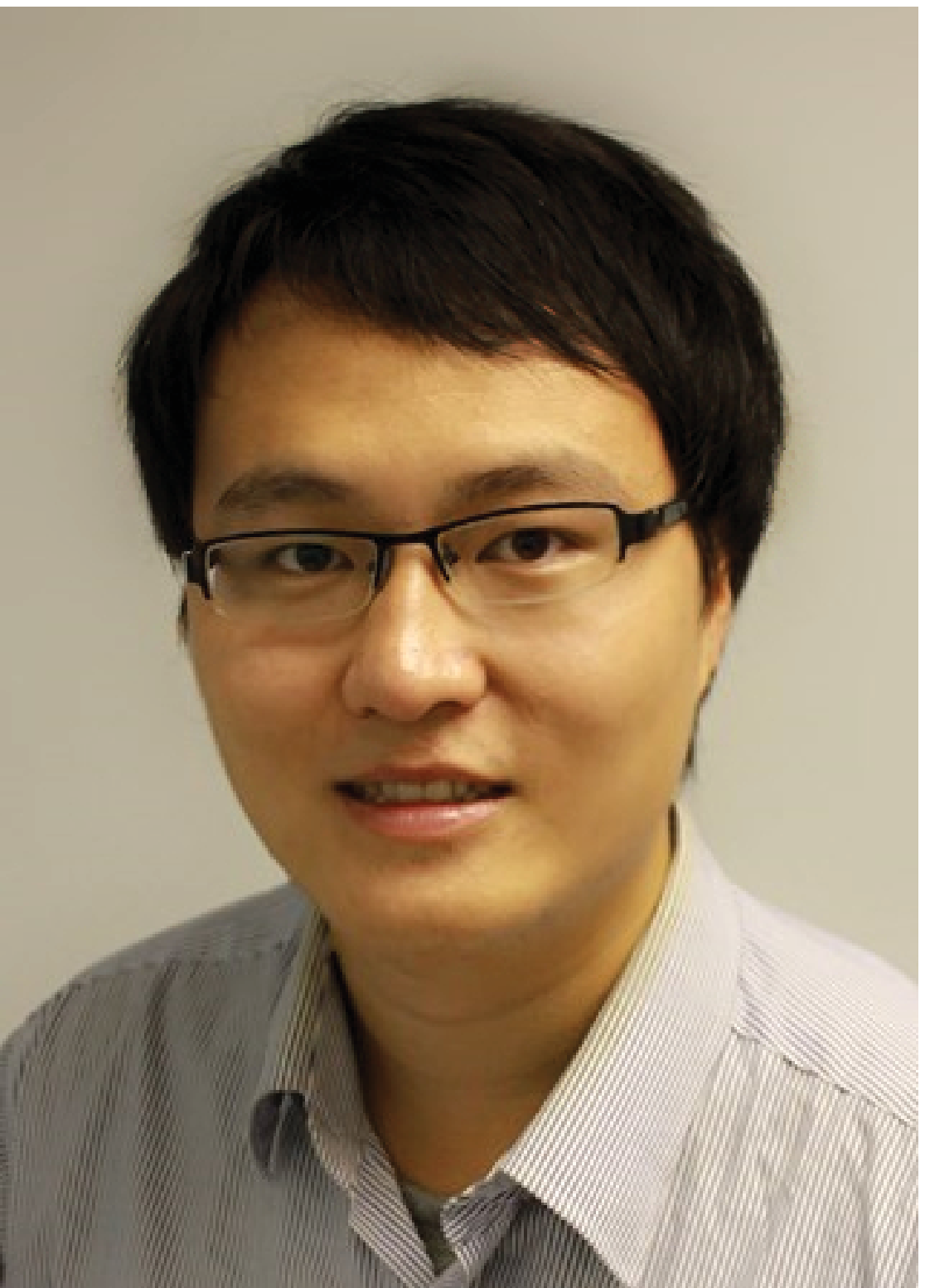}}]{Suzhi Bi}
(S'10-M'14) received his Ph.D. degree in Information Engineering from The Chinese University of Hong Kong, Hong Kong in 2013. He received the B.Eng. degree in communications engineering from Zhejiang University, Hangzhou, China, in 2009. He is currently a research fellow in the Department of Electrical and Computer Engineering, National University of Singapore, Singapore. From June to August 2010, he was a research engineer intern at Institute for Infocomm Research (I$^2$R), Singapore. He was a visiting student in the EDGE lab of Princeton University in the summer of 2012. His current research interests include MIMO signal processing, medium access control in wireless networks and smart power grid communications. He is a co-recipient of Best Paper Award of IEEE SmartGridComm 2013.
\end{IEEEbiography}

\begin{IEEEbiography}
[{\includegraphics[width=1in,height=1.25in,clip,keepaspectratio]{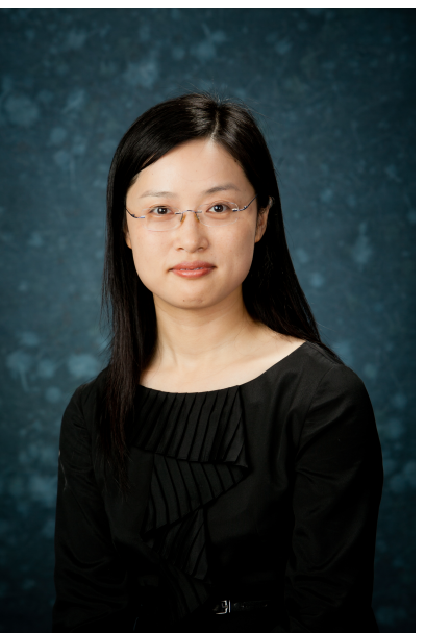}}]{Ying Jun (Angela) Zhang} (S'00-M'05-SM'11) received her Ph.D. degree in Electrical and Electronic Engineering from the Hong Kong University of Science and Technology, Hong Kong in 2004. She received a B.Eng in Electronic Engineering from Fudan University, Shanghai, China in 2000.

Since 2005, she has been with Department of Information Engineering, The Chinese University of Hong Kong, where she is currently an Associate Professor. She was with Wireless Communications and Network Science Laboratory at Massachusetts Institute of Technology (MIT) during the summers of 2007 and 2009. Her current research topics include resource allocation, convex and non-convex optimization for wireless systems, stochastic optimization, cognitive networks, MIMO systems, etc.

Prof. Zhang is an Executive Editor of IEEE Transactions on Wireless Communications, an Associate Editor of IEEE Transactions on Communications, and an Associate Editor of Wiley Security and Communications Networks Journal. She was a Guest Editor of a Feature Topic in IEEE Communications Magazine. She has served as a Workshop Chair of IEEE ICCC 2013 and 2014, TPC Vice-Chair of Wireless Communications Track of IEEE CCNC 2013, TPC Co-Chair of Wireless Communications Symposium of IEEE GLOBECOM 2012, Publication Chair of IEEE TTM 2011, TPC Co-Chair of Communication Theory Symposium of IEEE ICC 2009, Track Chair of ICCCN 2007, and Publicity Chair of IEEE MASS 2007. She was a Co-Chair of IEEE ComSoc Multimedia Communications Technical Committee, an IEEE Technical Activity Board GOLD Representative, IEEE Communication Society GOLD Coordinator, and a Member of IEEE Communication Society Member Relations Council (MRC). She is a co-recipient of 2011 IEEE Marconi Prize Paper Award on Wireless Communications, and a co-recipient of Best Paper Award of IEEE SmartGridComm 2013. As the only winner from Engineering Science, she has won the Hong Kong Young Scientist Award 2006, conferred by the Hong Kong Institution of Science.

\end{IEEEbiography}

\end{document}